\documentclass[12pt,a4paper]{article}

\newcommand{\rmd}{\mathrm{d}}
\newlength{\abstwidth}
\newenvironment{Itemize}{\begin{list}{$\bullet$}%
{\setlength{\topsep}{0.2mm}\setlength{\partopsep}{0.2mm}%
\setlength{\itemsep}{0.2mm}\setlength{\parsep}{0.2mm}}}%
{\end{list}}
\newcounter{enumct}

\begin{document}
%set sloppy attitude to line breaks
\sloppy

\begin{flushright}
LU TP 13-34\\
MCnet-13-13\\
September 2013
\end{flushright}

\vspace{\fill}

\begin{center}
{\LARGE\bf Challenges for QCD theory\\[2mm] 
-- some personal reflections --}\\[10mm]
{\Large T.~Sj\"ostrand} \\[3mm]
{\it Theoretical High Energy Physics,}\\[1mm]
{\it Department of Astronomy and Theoretical Physics,}\\[1mm]
{\it Lund University,}\\[1mm]
{\it S\"olvegatan 14A,}\\[1mm]
{\it SE-223 62 Lund, Sweden}
\end{center}

\vspace{\fill}

\begin{center}
\begin{minipage}{\abstwidth}
{\bf Abstract}\\[2ex]
At the LHC \textit{all} processes are QCD ones, whether ``signal'' or 
``back\-ground''. In this review the frontiers of current QCD research
are addressed, towards increased understanding, improved calculational
precision, and role in potential future discoveries. Issues raised include\\ 
the limits of perturbative QCD calculations and parton distribution
usage,\\ 
the nature of multiparton interactions,\\ 
the impact of colour reconnection on physical observables,\\ 
the need for progress on hadronization modelling,\\ 
the improvements of parton showers and their combination with the
matrix-element description,\\ 
the use of QCD concepts in Beyond-the-Standard-Model scenarios, and\\ 
the key position of event generators and other software in the successful 
exploration of LHC physics.\\ 
On the way, several questions are posed, where 
further studies are needed. 
\end{minipage}
\end{center}

\vspace{\fill}
\noindent\rule{60mm}{0.3mm}\\
{\footnotesize To appear in the proceedings of the Nobel Symposium on LHC 
results, Uppsala, Sweden, 13 -- 17 May 2013}

\clearpage

\section{Introduction}

Given that LHC collides hadrons, it follows that \textit{all} processes
are QCD ones. The calculation of even the most exotic process is done
within a perturbative QCD framework, with parton distributions and 
higher-order QCD corrections as important ingredients, and with other
QCD-calculated processes as background, often with normal QCD jets 
production on top of the list.

Today the QCD Lagrangian is well tested, and is not an issue.
Nevertheless challenges abound, and here we collect them into three
partly overlapping frontiers, as always with the proviso that 
ultimately everything hangs together.
\begin{Itemize}
\item Understanding: many established phenomena still lack a proper
theoretical description, such as confinement, the quark--gluon plasma,
the hadronization process, the behaviour of interactions in the 
small-$x$ limit, multiparton interactions, and colour reconnection.
\item Precision: higher-order matrix elements and parton distributions
allow for higher precision, but loop calculations are demanding and
progress takes time. Parton showers offers a complementary approach
that is convenient in collinear and soft regions,  but the matching 
between the two descriptions is nontrivial.
\item Discovery: characterizing signal and background properties
is essential for searches, notably when jets are produced. In addition, 
several scenarios for BSM physics involve new aspects where QCD offers 
a template.    
\end{Itemize}
Examples from overlapping regions is that jet properties and the
proton spin involve both understanding and precision,  that the 
higher-order calculation of BSM involve both precision and discovery,
and that the mass definition of coloured particles (like the top)
involve both understanding and discovery.

In the following I will touch on several of these topics. It is 
beyond the scope of this brief presentation to cover all interesting
issues, so what follows is a subjective selection, with subjective 
opinions. Some of the topics not discussed here are covered in the 
experimental QCD presentation of A. De Roeck \cite{deRoeck}. 
Another useful reference is the recent minireview by G. Salam 
\cite{salamRev}. 

\section{Perturbative QCD and parton distributions}

There is a steady stream of new calculations being presented, and  
previous wish lists of NLO calculations have been checked off.
All this is thanks to a healthy, strong community of calculators,
even with an influx from the superstring side, that also increasingly 
make new results available in the form of public codes.
More specifically
\begin{Itemize}
\item LO calculations are fully automatized up to the order of six
to eight final-state particles, the limit being set by computer 
resources more than anything else.
\item NLO calculations are also in the process of being automatized.
Currently the limit is somewhere around four final-state particles.
\item NNLO is the current calculational frontier, with only
one-particle processes fully under control, but two-body final
states are now in the process of being mastered.
\item Quite apart from the matrix elements themselves, the
phase-space sampling can be a bottleneck, and efficient methods
are needed to speed up calculations.
\end{Itemize}

Among all the many calculations, maybe the gg~$\to$H$^0$ NNLO one 
\cite{sigmaHiggs} 
deserves special attention. It showed that the convergence of the
cross section is slow, going from LO to NLO to NNLO, and also in
terms of the width of the scale variation band. This is interesting
to understand, and also is highly relevant for tests of the 
Standard-Model nature of the Higgs. The results of several other 
calculations are used in other presentations at this symposium. 

Also when it comes to parton distributions there exists a healthy 
competition/collaboration between a few different groups, regularly
providing new tunes to available data \cite{thorneRev}.
High precision has been obtained with NLO fits, and there are some 
first NNLO fits. Needless to say, these sets are to be combined with 
the above NLO/NNLO calculations.

The NLO framework tends to break down if the description is extended
to low $Q^2$ scales, say below 4~GeV$^2$. Data in combination with
the NLO splitting kernels tends to drive the gluon negative (or at 
least very close to zero) in the low-$x$ region, a behaviour that 
becomes even more marked at NNLO. More generally, the NLO and NNLO 
frameworks are unstable at low $Q^2$. Interestingly, it appears that 
resummed PDFs recover the physical LO behaviour. Thus\\
\textit{Open question 1: would it be possible to develop a new 
calculational scheme, wherein both MEs and PDFs are systematically
resummed to increasing order, in such a way that both retain a
cleaner physical interpretation than the traditional 
$\overline{\mathrm{MS}}$ route offers?}

\section{Multiparton interactions and colour reconnection}

Given the composite nature of the incoming protons, it is inevitable 
that multiparton interactions (MPIs) play an important role.
Indeed, in most models they are the driving force for the structure
both of minimum-bias and underlying events. The most direct
manifestation of MPIs is the very long tail out to high multiplicities
in minimum-bias events, where most of the particles have no apparent 
association with hard jets. At the other end, studies that involve 
correlations between four hard jets (or three jet and a photons, 
or two jets and a weak gauge boson)  corroborate the MPI picture, but 
only probe a tiny fraction of its total cross section \cite{mpiData}.

The dominant QCD processes involve $t$-channel gluon exchange, which
leads to a $\rmd p_{\perp}^2 / p_{\perp}^4$ divergence for 
$p_{\perp} \to 0$. This behaviour must be regularized, e.g. by a 
dampening to $\rmd p_{\perp}^2 / (p_{\perp 0}^2 + p_{\perp}^2)^2$.
The obvious scale would have been 
$p_{\perp 0} \sim \Lambda_{\mathrm{QCD}} \sim 0.3$~GeV,
but empirically a value like $p_{\perp 0} \sim 2 - 3$~GeV is called for
\cite{mpiTS}. This raises\\ 
\textit{Open question 2: is the size of $p_{\perp 0}$ set by colour 
screening effects inside the proton and, if so, how could it be
calculated rather than fitted?}

While MPIs produce outgoing partons, these need to hadronize.
As will be discussed later, in the Lund string model quarks and 
antiquarks sits at the ends of strings while gluons form kinks on
the string, and it is these strings that fragment to produce the 
primary hadrons. The way the strings are stretched is based on the 
colour assignment. The issue is then the reliability of the naive 
perturbative assignments, specifically whether colours could be
rearranged before the hadronization stage. One early example of this is 
J/$\psi$ production in B meson decay \cite{fritzsch}, where the colour 
singlet nature of the W puts the c and $\overline{\mathrm{c}}$ in 
separate singlets in b$\to$cW$\to$c$\overline{\mathrm{c}}$s decays. 

W pair production at LEP~2 offered a interesting test bed for such
concepts, i.e. whether the q$\overline{\mathrm{q}}$ pair produced in 
each W decay would hadronize separately or whether e.g.\ the q
from one W could hadronize together with the $\overline{\mathrm{q}}$
of the other. Notably, this could mess up W mass determinations.
Unfortunately, results were not conclusive.
\begin{Itemize} 
\item Perturbative effects are suppressed for a number of reasons,
notably that hard-gluon exchanges would force the W propagators
off-shell, giving a negligible uncertainty 
$\langle \delta M_{\mathrm{W}} \rangle \leq 5$~MeV \cite{reconKhoze}.
\item Several nonperturbative colour reconnection models predicted 
large effects and could promptly be ruled out. The more conservative
ones \cite{reconKhoze} could not be excluded, although they were not 
favoured \cite{reconData}, and gave 
$\langle \delta M_{\mathrm{W}} \rangle \sim 40$~MeV.
\item Additionally Bose-Einstein effects, i.e.\ that the wave function
of identical integer-spin hadrons should be symmetrized, could 
affect the separate identities of the W$^+$ and $W^-$ decay products.
Effects on $\langle \delta M_{\mathrm{W}} \rangle$ could be as large as
100~MeV, but again more likely around 40~MeV \cite{beLonnblad}. 
An effect of the latter magnitude is disfavoured by data, but again 
not fully ruled out \cite{beData}. 
\end{Itemize}

Hadron collisions offers a much more busy environment that did LEP~2,
however. A typical LHC collision may involve five MPIs. The dominant
gg$\to$gg processes each pulls out a colour octet from the two
incoming beams, thereby naively leading to two (triplet) strings
being stretched between the two beam remnants. Since the transverse 
size of these strings is the same as that of proton this leads to 
ten string almost on top of each other over much of the rapidity range.
It would be surprising indeed if this did not have consequences.  

The most direct probe of such effects is how the average transverse
momentum of charged particles varies with the charged multiplicity.
In cases where each subcollision system fragments independently
one would expect $\langle p_{\perp} \rangle (n_{\mathrm{ch}})$ to be
essentially flat --- the variation in multiplicity mainly reflects
the variation in the number of MPIs, but a higher multiplicity just
means more of the same. 

In reality, it is not realistic to assume that the beam remnants
acquire an arbitrarily large colour charge. This will naturally 
connect several interactions, such that strings are not pulled all
the way out to the remnants. The $p_{\perp}$ kicks from the MPIs 
themselves thus gets to be shared between fewer hadrons, and 
$\langle p_{\perp} \rangle (n_{\mathrm{ch}})$ obtains a rising trend.
This is nowhere near, however, and models require a significant
amount of reconnections, wherein partons from different MPIs get
their colours dramatically exchanged, in such a way that the total
string length is reduced \cite{mpiTS}. Then the hadronic multiplicity 
increases slower-than-linear with the number of MPIs. 
That way, it is possible to obtain a good description of data, but\\ 
\textit{Open question 3: what physics mechanisms are at play when
several colour fields overlap and how should they be modelled
correctly? }

Recently it has also been noted that colour reconnection in pp
can give some of the observed effects similar to the collective flow 
of heavy-ion collisions \cite{collectflow}.

\section{The mass of coloured unstable particles}

The top quark, as well as the W and Z gauge bosons, travel a distance
$c \tau \approx 0.1$~fm before they decay, i.e.\ significantly less than
a proton radius. Therefore their decay takes place right in the middle
of the hadronization region, and so quarks (and gluons) produced in the 
decays are subject to the reconnection issues already discussed above.
(By contrast the Higgs is so long-lived, $c \tau \approx 50$~fm, that 
there is no problem.)   

Current top mass measurements at the Tevatron and the LHC now have 
statistical errors of the order 0.5 GeV, and quote systematic errors
below 1 GeV \cite{topData}. These measurements heavily rely on comparisons 
with event generators. What is quoted as the top mass is actually the 
mass parameter used in the generators, which is close to the pole mass,
but not necessarily identical. It is also to be assumed that the 
handling of higher-order matrix-element and parton-shower corrections
is under control. Colour reconnection uncertainties then come on top 
of that. Model studies have suggested a total (perturbative + 
nonperturbative) uncertainty approaching 1~GeV \cite{topSkands}, 
almost saturating the current systematic-error budget. 

Clearly this issue needs to be studied further, to try to constrain the
possible magnitude of effects from data itself. Effects of
colour reconnection should have a dependence on the event kinematics,
which would allow to test and constrain models. Such studies have 
already begun in CMS \cite{topStudCMS}, although statistics does not yet 
allow any conclusions to be drawn. In view of the long and winding path
ahead of us, one may look for alternatives:\\
\textit{Open question 4: is it possible to find better 
(theoretical + experimental) mass definitions for coloured unstable
particles?}

\section{Hadronization}

The oldest hadronization model still in common use is the Lund string one
\cite{lundString}.
When a colour-singlet q$\overline{\mathrm{q}}$ pair is pulled apart,
it is assumed that the colour field lines are pulled together to a tube-like
region, giving a transverse radius $\sim$0.7 fm comparable with the 
proton one, and a linear confinement potential $V(r) \approx \kappa r$,
$\kappa \approx 1$~GeV/fm.

The string does not get very long, however, since it breaks by the 
production of new q$\overline{\mathrm{q}}$ pairs inside the string,
that screen the colour charges of the endpoint. Repeated such breaks 
give rise to the primary hadrons, which are distributed approximately
flat in rapidity space, but with short-range anticorrelations from
local energy (and flavour) conservation in the string, and longer-range 
effects from global conservation. The required tunneling of quarks 
with nonzero transverse mass gives a Gaussian transverse momentum 
spectrum and a suppression of heavier quarks (and hadrons). While generally 
supported by LEP data, its weak point is that it relies on a number of 
parameters for the flavour production properties.

Multiparton configurations are considered in the $N_c \to \infty$ limit,
so that all colours are unique and gluons carry a separate colour and 
anticolour index. This means that a string is stretched from a quark 
end via a number of gluons to an antiquark end (or in a closed gluon 
loop). Diquarks are treated like antiquarks, to first approximation. 
LEP q$\overline{\mathrm{q}}$g events thus contain two string pieces,
one from q to g and another from g to $\overline{\mathrm{q}}$. 
Each of those two pieces can be viewed as boosted copies of the simple
q$\overline{\mathrm{q}}$ string, tied together at the gluon corner.
No new parameters are needed. The model predicts a depletion of 
particle production in the angular region between the q and   
$\overline{\mathrm{q}}$ where there is no string, well verified 
in data.

The main alternative is cluster models \cite{cluster}, wherein the 
parton shower to a low cutoff scale is complemented by final 
g$\to$q$\overline{\mathrm{q}}$ branchings that splits the system
into smaller singlets. Originally these were assumed to decay 
isotropically, but for larger singlets a preferred decay direction
is introduced along string ideas.

Both the string and cluster models were developed in an e$^+$e$^-$
environment and then applied to pp/p$\overline{\mathrm{p}}$ events
with moderate extensions. Notably the large string (or cluster) 
overlaps at the LHC, already described above, are tackled by 
mechanisms like colour reconnection that do not put in question
the relevance of the string model as such. There have been some studies
of fields in higher colour representations (``colour ropes'') \cite{ropes},    
but rather little else that attempts to bridge the gap between 
the simple e$^+$e$^-$ picture and a full-blown quark--gluon plasma.
Thus\\
\textit{Open question 5: can one develop new hadronization models 
more relevant for the busy LHC environment, while still not clashing
with established e$^+$e$^-$ phenomenology in that limit?}

\section{Parton showers}

Traditional showers are constructed by combining repeated $1 \to 2$ 
branchings, $a \to b c$. Partons originally assumed massless may 
need to be assigned virtualities in the process, and thus local or 
global correction procedures need to be introduced to handle
energy--momentum conservation. Emissions can be ordered in
angle, virtuality or $p_{\perp}$, with restrictions from colour
coherence phenomena \cite{webberRev,evtGenRev}. 

An alternative is the dipole shower \cite{lundDip}, inspired by 
the Lund string and the St.~Petersburg dipole \cite{stpDip}. In it 
branchings are instead of the $2 \to 3$ character, $a b \to c d e$. 
When $a$, $c$ and $d$ are close to collinear (and flavours match) 
the $a$ can be viewed as the radiator and the $b$ as a recoiler, 
there to ensure local energy--momentum conservation, 
$p_c + p_d + p_e = p_a + p_b$. Again relying on the 
$N_C \to \infty$ limit, consecutive emissions 
give rise to an increasing number of separate dipoles, e.g.\ an
original q$\overline{\mathrm{q}}$ dipole after a gluon emission turns
into two dipoles, qg and g$\overline{\mathrm{q}}$, essentially smaller
replicas in their respective rest frames. When emissions are ordered 
in $p_{\perp}$ also coherence conditions are fulfilled.

Nowadays the dipole-style showers are the ones most commonly used.
From their origin in final-state radiation they have been extended
to initial-state radiation, and in the form of Catani--Seymour
dipoles \cite{catSey} found a use also in NLO calculations.

A crucial ingredient of the shower approach is the Sudakov form factor
$\Delta(p_{\perp 1}^2, p_{\perp 2}^2)$, which expresses the no-emission 
probability between the $p_{\perp 1}$ and $p_{\perp 2}$ evolution scales.
It can be obtained by an appropriate exponentiation of the real-emission 
probability over that range, and thus ensures that emission probabilities
never exceed unity. 

The universal nature of showers makes them very convenient to add onto 
a fixed-order calculation, to construct more realistic final states.
Still the shower approach faces challenges, and work is ongoing in
different directions \cite{newShowers}, e.g.\ the following ones. 
\begin{Itemize}
\item The accuracy is formally only to LL, even if the many beyond-LL 
aspects added to the showers should include the bulk of NLL effects
(energy--momentum conservation, $\alpha_{\mathrm{s}}(p_{\perp}^2)$,
full $z$ dependence of splitting kernels, coherence, \ldots).
\item Most showers do not cover the full phase space, but 
leave some gaps. 
\item Nonleading colour terms are neglected when defining the colour
flow (but not in the splitting kernels). 
\item It may become relevant to include weak gauge boson emission, 
notably for high-$p_{\perp}$ jets at the LHC. 
\item Most importantly, showers should attach as well as possible to 
the matrix elements that they are combined with, which is the topic 
of the next section. 
\end{Itemize}

\section{Matching/merging of matrix elements and parton showers}

In some respects the improved calculational capability for matrix
elements has reduced the need for parton showers. Today it would be 
possible to describe an 8-jet final state purely by matrix elements,
which would have been infeasible ten years ago. In other respects,
however, parton showers are as needed now as ever. One first point 
is that the perturbative description must go down to scales 
of the order of 1 GeV, where nonperturbative hadronization can take 
over, and at such low scales the effective number of partons can 
exceed any matrix-elements capacity. Other points will be raised
as we go along.

Given the need both for the more precise ME description and the 
more flexible PS one, quite some work has gone into the best ways to 
combine the two. Such efforts go under the name of matching or merging;
with a distinction that largely is author-dependent. For many years 
this work has either been for multiple legs, i.e.\ final-state partons,
or more loops, i.e.\ NLO. Currently the emphasis has shifted to combining 
the two, i.e. to have multileg at NLO. 

The key point about LO MEs is that they are inclusive: a $2 \to n$ 
calculation gives the rate of having at least $n$ partons, but with 
no upper limit. This means that the observable exclusive $n$-parton
rate is not directly related to the LO ME calculation. It is here the 
Sudakov form factor enters: by encoding the no-emission probability
it can be used to turn an inclusive calculation into an exclusive one.    

This is the basic idea of the CKKW approach \cite{ckkw}: use MEs for 
real emissions and Sudakovs for the virtual corrections needed to obtain 
an exclusive picture. To calculate those Sudakovs, it is necessary to 
construct a fictitious shower history, such that no-emission probabilities 
can be calculated also for intermediate propagators. When several 
histories are possible, their relative probabilities are used to 
pick one. With a shower history at hand, it also becomes possible to
reweight events originally picked with a fixed $\alpha_{\mathrm{s}}$
(necessary to preserve gauge invariance) to have a running
$\alpha_{\mathrm{s}}(p_{\perp}^2)$ at the $p_{\perp}$ scale of each 
branching.

The original CKKW scheme was based on analytic Sudakovs, which are
rather crude, and this is no longer used. Instead the CKKW--L approach 
is based on using trial showers to provide the Sudakov factors 
\cite{ckkwl}. For instance, given a $p_{\perp}$-ordered shower algorithm, 
a $p_{\perp}$-ordered history is constructed from the ME information.
Then, for each step from $n$ to $n+1$ partons, a trial shower is
started up from the $n$-parton topology at scale $p_{\perp n}$,
and if a branching occurs above $p_{\perp n+1}$ the event is rejected.
This way the Sudakov suppression includes exact kinematics, running
$\alpha_{\mathrm{s}}$, coherence effects, and so on. The more accurate 
the shower algorithm, the more trustworthy the Sudakov factors,
so the incentive for improved showers remains high. 

An alternative to CKKW--L is the MLM algorithm \cite{mlm}. 
Also here showers are used as a means to go from inclusive to 
exclusive event samples. Without going into the details, the difference 
is that MLM does not micromanage the shower, but only considers whether 
the final jet state after showers matches the original parton state. 

At the other frontier, matching with NLO calculations, there are two 
well established approaches, MC@NLO \cite{mcanlo} and POWHEG 
\cite{powheglund,powheg}. In retrospect these can be combined into 
one master formula, in simplified form (for a fixed Born-level topology)
\[
\rmd\sigma = \rmd\sigma_{R,\mathrm{hard}} 
+ (\sigma_B + \sigma_{R,\mathrm{soft}} + \sigma_V) 
\left[ \frac{\rmd\sigma_{R,\mathrm{soft}}}{\sigma_B} 
\exp \left( - \int \frac{\rmd\sigma_{R,\mathrm{soft}}}{\sigma_B} 
\right) \right]
\]
where $\sigma_B$ is the $n$-body Born term, 
$\rmd\sigma_R = \rmd\sigma_{R,\mathrm{hard}} + \rmd\sigma_{R,\mathrm{soft}}$
are the real-emission terms to $n+1$-body states, and
$\sigma_V$ are all virtual corrections (including PDF counterterms)
for the $n$-body states. The expression in square brackets is normalized 
to unity, and can be viewed as a parton-shower-like downwards evolution
in an emission hardness variable like $p_{\perp}$, with the exponential 
providing the Sudakov factor for not having a harder emission than the 
currently considered one. (For a nonzero lower $p_{\perp\mathrm{min}}$ 
cutoff there also appears an additional term inside the square bracket, 
to represent that this lower cutoff sometimes can be reached without 
an emission, and thereby unitarity is preserved.) The prefactor 
$(\sigma_B + \sigma_{R,\mathrm{soft}} + \sigma_V)$, 
with $\sigma_{R,\mathrm{soft}} = \int \rmd\sigma_{R,\mathrm{soft}}$, 
gives the cross section for this exponentiated part, whereas 
$\int \rmd\sigma_{R,\mathrm{hard}}$ gives the cross section for the 
non-exponentiated part. Thus all events contain an emission 
(apart from those falling below the $p_{\perp\mathrm{min}}$ cutoff).
A normal shower can take over below the $p_{\perp}$ scale of the one 
and only ME ``emission'' (or below $p_{\perp\mathrm{min}}$).   

In this formula, POWHEG corresponds to the special case
$\rmd\sigma_{R,\mathrm{hard}} = 0$, i.e. the whole cross section is
exponentiated. Specifically, the $n+1$-body high-$p_{\perp}$ tail is
multiplied by a $K = (\sigma_B + \int\rmd\sigma_R + \sigma_V) / \sigma_B$
factor typically above unity. This is formally a NNLO ambiguity
and so allowed in an NLO approach, but is not appreciated by all.

The rival MC@NLO is based on having a shower that attaches well 
to the $\rmd\sigma_R$ behaviour in the $p_{\perp} \to 0$ limit, 
Sudakov factor uncounted, as a shower should. It is this 
shower-without-Sudakov rate that is associated with 
$\rmd\sigma_{R,\mathrm{soft}}$. Furthermore, by picking a ``bad'' 
shower algorithm, that falls off faster than $\rmd\sigma_R$ at 
large $p_{\perp}$, one ensures that the unexponentiated 
$\rmd\sigma_{R,\mathrm{hard}}$ dominates in this region.
This term is not multiplied by a $K$ factor, which some people prefer.
The price to pay is a stronger bond to a specific shower, and the
possibility of negative-weight events in regions where the shower
overestimates the matrix elements.

In the few cases where the NNLO answer is available, such as 
Higgs production, it turns out that the $K$-factor rescaling of 
POWHEG gives a more accurate $p_{\perp}$ spectrum than MC@NLO 
\cite{phhiggs}. That it, whatever physics causes the large $K$ 
factor of gg$\to$H also seems to give a correspondingly large 
correction to gg$\to$Hg.

As already mentioned, the current frontline is to combine the 
multileg matching technique with NLO input. Taking the Higgs case 
as example, there are two reasons for this. Firstly the multileg
matching does not offer a total Higgs cross section more accurate 
than LO. Secondly, the basic NLO scheme only offers NLO for 
the total Higgs rate; it is LO for H + 1 jet, and gives nothing 
beyond that. Or, alternatively, NLO for H + 1 jet and LO for for 
H + 2 jet, if one starts one order up. Current technology now
can handle both H and  H + 1 jet to NLO, and more jets to LO
\cite{shmatch,hwmatch,pymatch}. It does not come without a price,
however, either of allowing spurious NNLO terms, or of having quite 
complicated formulae, with negative-weight events needed to preserve 
the normalizations. Recently there has even been a first implementation
that preserves the NNLO total cross section for Higgs production
\cite{mimatch}, again at the price of significant complexity. So\\
\textit{Open question 6: is it possible to construct a generic, 
transparent, robust and reliable approach to matching beyond LO?}\\
(Note that Sudakov factors and resummation are related to each other,
so progress on open question 1 could go a long way in this direction,
but we should not wait for that to happen.)

\section{Event generators}

LHC events are of daunting complexity, if one starts to consider 
all the different physics mechanisms that are at play in them.
In this article  we have mentioned matrix elements and parton
distributions, multiple partonic interactions, initial- and 
final-state showers, beam remnants, colour reconnection,
hadronization, decays and Bose-Einstein effects. Further
mechanisms and aspects have been proposed, and new ideas may
still come along.

Currently the only known way to address this complexity is through 
event generators \cite{evtGenRev}, where the overall task is broken 
into more manageable subtasks, along the lines of the above list.
Monte Carlo methods are used to represent quantum mechanical choices
at all steps along the way. Given all the limitations, it is fair 
to ask\\
\textit{Open question 7: can one find better alternatives to event 
generators, that have a corresponding breadth of applicability?}

The three workhorses for LHC pp physics are \textsc{Herwig} \cite{hwMan}, 
\textsc{Pythia} \cite{pyMan} and \textsc{Sherpa} \cite{shMan}. 
Since they set out to describe the same physical reality, they do share 
many common traits.Nevertheless there are distinguishing features,
and different historical roots, reflecting the topics of interest 
at the time.
\begin{Itemize}
\item \textsc{Pythia} has it roots in the Lund string code
begun in 1978, and has retained a high profile in soft physics, 
such as multiparton interactions.
\item \textsc{Herwig} originated in 1984 from the introduction 
of coherent shower evolution through angular ordering, and this
has remained the hallmark of the program. 
\item \textsc{Sherpa} dates back to 2000 and has in particular 
been developed to handle CKKW and related kinds of ME/PS matching
procedures.
\end{Itemize}
The (EU-funded) MCnet \cite{mcnet} offers common activities of 
these collaborations, and other related projects, such as summer 
schools on event generator physics.

Since the generators involve many parameters, mostly related to 
nonperturbative physics, there is a need to tune them to data. 
Once tuned, they can then be applied to make predictions for 
observables not yet studied. The key assumption is that the
generators contain the correct physics, and that therefore good
tunes can be found. Counterexamples may already be at hand, e.g.\
in terms of a somewhat different flavour composition at LEP and LHC.
The tuning effort is shared between the generator authors, 
the experimental collaborations, and some separate efforts 
\cite{tunes}. Much data, but far from all that would have been useful, 
is made available in such a form that it can be compared with the 
output from generators \cite{rivet}, so there is room for improvement. 

In addition to the above three generators, a plethora of more specialized
programs exist, and new ones are added all the time.
These include complete generators for QCD
physics (heavy ions, cosmic rays), separate shower programs,
matrix-element generators and ditto libraries, Feynman rule 
generators, PDF libraries, specialized programs for BSM scenarios 
(mass spectra, decays, matrix elements, \ldots), jet finders 
(including jet grooming techniques) and
other analysis packages (including detector simulation and 
parameter constraints from data). Some are projects at the same 
scale as the three standard generators, such as MadGraph \cite{madgraph},
and \textsc{Geant} \cite{geant} of course is orders of magnitude bigger. 
The most impressive point is that all of these different kinds of 
software can come together to produce meaningful simulations of 
LHC physics. One reason that this has been possible is that
common standards have been developed for a number of interfacing
tasks. 

\section{QCD and BSM physics}

Many/most scenarios for BSM involve coloured particles, and so 
QCD is not only a matter of production processes but also of the 
consecutive fate. When the coloured particles are short-lived and 
decay to standard particles this description need not involve much 
more than the kind of framework already developed e.g.\ to handle 
showers in top decay. But there are cases that go beyond the simple 
scenarios, such as the following four.
\begin{Itemize}
\item Baryon number violation is allowed in some SUSY scenarios.
If a neutralino is the lightest supersymmetric particle it can decay 
to three quarks. On the shower level the standard radiator--recoiler
picture has to be extended, and on the hadronization level 
the fragmenting string has a Y-shaped topology with a junction 
in the middle \cite{bnv}.
\item Other SUSY scenarios allow for long-lived squarks or gluinos,
that then have time to fragment into so-called $R$-hadrons.
The squark will be at the end of a string, and so hadronization is
not so different from that of a heavy quark, but the gluino will
be a massive kink inside a string, which has no precedent in the SM.
In addition to the possible formation of ``mesons'' 
$\tilde{\mathrm{g}}$q$\overline{\mathrm{q}}$ and ``baryons''
$\tilde{\mathrm{g}}$qqq, also ``glueballs'' $\tilde{\mathrm{g}}$g
could be allowed \cite{rhad}.
\item If black holes can be formed, notably in scenarios with extra 
dimensions, they will rapidly evaporate by the emission of all kinds 
of particles, but mainly by hadrons. The emission properties depends
on the temperature of the black hole, which increases as its mass
drops, requiring an evolution and hadronization approach quite 
different from the normal one \cite{bhole}.  
\item Hidden-valley scenarios could allow for a repetition of a
strong-interaction frame\-work in some secluded sector, with showers
and hadronization. Seepage back into the normal sector would partly
reveal the pattern, which therefore needs to be modelled. It is
also possible to have particles with both SM and hidden charges,
where thus radiation into the two sectors may be interleaved \cite{hval}.  
\end{Itemize}  

\section{Summary and outlook}

This (biased) selection of topics illustrates the breadth of current
QCD-based research, both in its own right and in support of all other
LHC physics studies. 

There are many other topics that would deserve attention, such as
\begin{Itemize}
\item Jet production rates, jet properties and jet algorithms.
\item Production of other SM (and BSM) particles, such as photons,
weak gauge bosons, quarkonia, top and the Higgs.
\item Identified particle production, such as the $\pi$/K/p composition.
\item Heavy-flavour (c and b) production topologies, especially
by shower evolution. 
\item Flavour production asymmetries, observed in baryon number transport
or B hadron composition, reflecting the beam proton valence flavours. 
\item Relations between minimum-bias and underlying-event physics. 
\item Total, elastic and diffractive cross sections.
\item Diffractive and forward physics.
\item Tests of small-$x$ evolution. 
\item The proton wave function and spin physics.
\item The ridge effect, signs of collective flow and other 
connections between pp and heavy-ion physics. 
\end{Itemize}
Fortunately many of these topics are brought up in other presentations 
at this symposium.

The overall picture is that the QCD community has been quite successful
in providing useful input to everybody working with LHC physics, 
from NLO/NNLO calculations to complete event generation. At the same
time, less emphasis is put on QCD for its own sake, partly because
nobody today would question the validity of QCD as such, partly because
true progress in the understanding of QCD will be very tough. The old
battle cry of ``solving QCD'' (be it by lattice or superstring methods)
seems as remote as ever, but that does not mean we should not try   
to do better than we can today. The seven questions in this presentation
are examples of issues that at least should be considered.

\section*{Acknowledgements}
Work supported in part by the Swedish Research Council, contract number
621-2010-3326.

\end{document}